\documentclass[twocolumn,secnumarabic,amssymb, nobibnotes, aps, prd]{revtex4-1}

\usepackage{graphicx}
\usepackage{subcaption}
\usepackage{caption}
\usepackage[]{siunitx}
\usepackage{mathptmx}
\usepackage{amsmath}
\usepackage{bm}
\usepackage{nicefrac}

\newif\ifcmtr
\cmtrtrue
\ifcmtr 
\newcommand{\cmtr}[1]{ %
   [\color{red} \textbf{#1} \normalcolor]%
}%
\else
\newcommand{\cmtr}[1]{ %
}%
\fi

\def\narrowfigurewidth{0.6\columnwidth}

\begin{document}

\title{Macroscopic equivalence for microscopic motion in a turbulence driven three-dimensional self-assembly reactor}%

\author{T.A.G. Hageman}
\affiliation{KIST Europe, Campus E7.1, 66125 Saarbr\"ucken, Germany}
\affiliation{University of Twente, PO Box 217, 7500 AE Enschede, The Netherlands}
\affiliation{Saarland University, 66123 Saarbr\"ucken, Germany}

\author{P.A. L\"othman}
\affiliation{KIST Europe, Campus E7.1, 66125 Saarbr\"ucken, Germany}
\affiliation{University of Twente, PO Box 217, 7500 AE Enschede, The Netherlands}
\affiliation{Saarland University, 66123 Saarbr\"ucken, Germany}

\author{M. Dirnberger}
\affiliation{Max Planck institute for Informatics, Campus E1.4, 66123 Saarbr\"ucken, Germany}
\affiliation{Saarland University, 66123 Saarbr\"ucken, Germany}

\author{M. Elwenspoek}
\affiliation{University of Twente, PO Box 217, 7500 AE Enschede, The Netherlands}

\author{A. Manz}
\affiliation{KIST Europe, Campus E7.1, 66125 Saarbr\"ucken, Germany}
\affiliation{Saarland University, 66123 Saarbr\"ucken, Germany}

\author{L. Abelmann}
\email[Email: ]{l.abelmann@kist-europe.de}
\affiliation{KIST Europe, Campus E7.1, 66125 Saarbr\"ucken, Germany}
\affiliation{University of Twente, PO Box 217, 7500 AE Enschede, The Netherlands}

\begin{abstract}
	We built and characterised a macroscopic self-assembly
  	reactor that agitates magnetic, centimeter-sized particles with a
  	turbulent water flow.  By scaling up the self-assembly processes to
 	the centimeter-scale, the characteristic time constant scale
  	also drastically increases. This makes the system a physical simulator of 
  	microscopic self-assembly, where the interaction of inserted particles 
  	are easily observable. Trajectory analysis of single particles reveals their
	velocity to be a Maxwell-Boltzmann distribution and it shows that their
	average squared displacement over time can be modelled by a confined
	random walk model,
	demonstrating a high level of similarity to Brownian motion. The
	interaction of two particles has been modelled and verified
	experimentally by observing the distance between two particles over
	time. The disturbing energy (analogue to
	temperature) that was obtained experimentally increases with sphere size,
	and differs by an order of magnitude between single-sphere and two-sphere 
	systems (approximately \SI{80}{\micro\joule} versus
	\SI{6.5}{\micro\joule}, respectively).
\end{abstract}

\date{8 August 2017}%
\maketitle

\section{Introduction}

Self-assembly is the process in which a disorganised system assembles in
a specific product without external interference. The final
properties of the assembly are determined by the properties of the
individual parts. Self-assembly is used extensively by nature; 
for example, in crystal growth, protein folding, the assembly of molecules 
into larger compounds, and the creation of complex organs such as the human brain.

Self-assembly is a prospective candidate for use in areas where conventional production
and assembly methods are problematic. Although it is not limited to specific 
dimensions~\cite{Whitesides2002}, self-assembly is especially applicable to
small scales~\cite{Elwenspoek2010}; for example, because conventional machining tools 
for three-dimensional construction are limited to larger feature sizes, while photo-lithography processes 
are two-dimensional in nature. Mastrangeli \emph{et al.'s}~\cite{Mastrangeli2009} review gives an excellent summary of this area,
ranging from nanosized DNA origami~\cite{Rothemund2006} to
magnetically folded milli-scale structures~\cite{Iwase2005}.

Arguably, one of the most promising applications will arise in the
semiconductor industry. As a result of the continuous downscaling of
fabrication processes, non-volatile data storage systems will at some
point run into its limit to store and process bits of information
using only a few atoms~\cite{Bennewitz2002}. To achieve higher data
densities, it is necessary to move to the third dimension. The first steps
in this direction have been taken by stacking wafers~\cite{Dellutri2006}
or layers~\cite{ Tanaka2007}. However, the stacking approach is not suitable to achieve truly three-dimensional
structures, in which both the resolution and extent of the features is
identical in all directions~\cite{Abelmann2010}. We believe that the most promising
production method is three-dimensional self-assembly.

Not only is three-dimensional self-assembly a prospective candidate for highly
repetitive memory structures, it will also open a path for more
complex electronics, such as processors. For instance, Gracias \emph{et
  al.}~\cite{Gracias2000} have designed
millimeter-sized polyhedra with integrated electronics. By
self-assembling these into crystals, functional electrical circuits have been demonstrated on a centimeter-scale. Scaling down the building blocks is a crucial
step towards scalability of the system as a whole.

It has been demonstrated that microscopic spherical particles can form regular 
structures upto centimeter-sized dimensions~\cite{Philipse1989}. By tuning the 
particle properties and/or the driving force of self-assembly, one can control the size and dimensions of the resulting structures~\cite{Manoharan2003,Rycenga2008}. 

Although major progress has been made in three-dimensional microscopic
self-assembly, observing the dynamic behaviour during the assembly
process remains a challenge due to the small size and time constants
involved. Several approaches have been explored to model and simulate these
processes \cite{Zhang2005,Grant2011,Whitelam2015}. However, these approaches rely on 
exhaustive Monte-Carlo simulations, scaling unfavourably with the number of
particles involved.
  
Magnetic forces have been used extensively as driving forces in self-assembly on 
all scales, together with various sources of agitating energy. 

When exposed to an external magnetic field, it has been demonstrated that 
nanoscopic magnetic rods form bundles (\cite{Love2003}) or multimers when driven by ultrasound (\cite{Tanaka2007}). 
Although paramagnetic spheres form chains, they will form ribbon structures 
(connected, parallel chains) for chains exceeding 30 particles (\cite{Darras2016, Messina2017}) and flower-like patterns result 
when magnetic and non-magnetic beads are mixed with ferrofluids (\cite{Erb2009}).
In the absence of an external magnetic field, a theoretical study of off-centered 
magnetic dipoles in spherical particles (\cite{Yener2016}) shows that lateral 
displacement of the dipoles results in structures that are more compact than chains. 
On millimeter-scales, magnetic forces and vibrations have been used to quickly 
and efficiently assemble particles with correct orientation on a template (\cite{Shetye2008, Shetye2010}). 
Also on centimeter-scales, magnetic forces have been used to form particles 
rather than structures, such as the spontaneously folding elastomeric sheets 
with embedded electronics; as demonstrated in (\cite{Boncheva2005}). 
Lash \emph{et al} (\cite{Lash2015}) showed that polystyrene beads self-assemble 
into HCP packed structures by solvent evaporation. Larger polystyrene particles ($>$\SI{18}{\micro\meter}) 
required additional disturbing energy (ultrasonic energy) as a disturbing energy source to self-assemble.
Macroscopic self-assembly processes on a centimeter scale are dominated by 
two-dimensional structures, where mechanical shaking is the most widely used source of disturbing energy.

Hacohen \emph{et 
al.}~\cite{Hacohen2015} demonstrated DNA-inspired patterned bricks with embedded 
magnets, self-assembling into a programmed structure, but report gravity bias. 
Stambaugh \emph{et al.}~\cite{Stambaugh2003} reported self-assembled 2D 
structures of centimeter-sized spherical particles with internal magnets 
that were shaken vertically, and observed different resulting structures that were based on 
particle concentration and magnet shape. Ilievsky \emph{et 
al.}~\cite{Ilievski2011} demonstrated self-assembly of centimeter-sized 
magnetic cubes into chains in a turbulent flow by submerging them in a rotating 
reactor filled with water, this way introducing eddy flows as a disturbing 
energy. They also introduced the concept of effective temperature, describing 
the motion of particles as if Brownian by nature. Even though the assembly process 
is three-dimensional, the resulting structures are limited to a single dimension and the dynamics involved are not studied. 

To build upon this work, we introduce an experimental setup, which is designated ``macroscopic self-assembly
reactor'', as a simulator for microscopic self-assembly. In this
reactor, we study the motion and interaction of centimeter-sized
objects. Particles are subject to a downward gravitational force and a drag force 
that is created by an upward water flow. We chose the particle density to balance 
these forces, causing them to appear weightless. Following Ilievski~\cite{Ilievski2011}, 
we use a turbulent water flow as an agitating source, simulating the Brownian motion on a microscopic scale. We employ permanent
magnets, resulting in attraction forces between the particles.

By increasing particle size from micrometers to centimetres,
not only the ease of observation but also the characteristic time
constants increase decidedly. This makes the self-assembly process visible
using conventional cameras. As a result of scaling up the system, the
environment also changes; laminar flows become turbulent while inertia effects become 
dominant. At the same time, Brownian motion becomes negligible. Therefore, it is crucial to study to what extent the
macroscopic system is a good simulator for microscopic environments,
which is the main topic of this publication.

\subsection{Organisation of this paper}
We characterise the motion and dynamics of particles in a
macroscopic self-assembly reactor. By observing the trajectories of a
single particle in the reactor, we quantify the similarity between
Brownian motion of said dynamics. By observing the interaction of two particles in the
reactor, we can characterise the most fundamental building block of the
self-assembly process, which is the interaction of magnetic spheres in a
turbulent environment. Section \ref{sec:theory} gives a
theoretical description of Brownian motion in a confined environment,
and provides a model of two-particle interaction based on
Maxwell-Boltzmann (M-B) statistics. Section \ref{sec:materials} 
introduces the reactor and magnetic particles in detail. Subsequently, 
in Section \ref{sec:results} we successfully analyse the extent to which 
the results of single- and two-particle experiments match our expectations based on the models.


\section{Theory}

\label{sec:theory}
Brownian motion is the apparent motion of microscopic particles suspended in a
fluid or gas, resulting from collisions with their surrounding
molecules, and it can be characterised by a three-dimensional random
walk. 

\subsection{Diffusion}
A random walk has an average
square displacement that increases linearly as time increases.
We can define a diffusion constant $D$ [\si{\square\meter\per\second}], which
in a system with three degrees of freedom links average displacement
$\left< x^2 \right>$ [\si{\square\meter}] to time $t$ [\si{\second}] according to

\begin{equation}
  \left< x^2 \right> = 6 D t.
  \label{eq:diffusion}
\end{equation}

This model holds only if the
average distance travelled is much smaller than the size of the
container in which the particles move. In our experiment this is not
the case and, therefore, container geometry needs to be taken into account.

To account for the confined space, we first consider a particle performing
a random walk along a single dimension. The particle displacement with
respect to its starting location after $t$ seconds is normally
distributed with variance $\sigma_x^2 = 2Dt$. Hence, the average squared
displacement $\left<x^2\right>$ is equal to the variance of the
distribution. The probability of the particle being outside of the confined space is zero. 
To account for this effect, we replace the normal distribution by a truncated normal distribution. If the truncation is symmetrical on both
tails of the normal distribution, $x_\text{t}$ [\si{\meter}], then the truncated
distribution is given by

\begin{equation}
  n_\text{t}(x,\sigma,x_\text{t}) = \begin{cases}
    \frac{n(x,\sigma)}{N(x_\text{t}, \sigma) - N(-x_\text{t}, \sigma)} & 
    -x_\text{t} \leq x \leq x_\text{t} \\
    0 & \text{otherwise},
  \end{cases}
\end{equation}

where $n(x, \sigma)$ is the normal distribution and $N(x, \sigma)$ is the 
cumulative normal distribution. The average squared displacement of a confined 
particle is the variance of this distribution:

\begin{equation}
	\label{eq:av_sq_disp_analytic}
	\left<x^2\right>  = \sigma^2 \left(1-\frac{x_\text{t} n(x_\text{t}, 
\sigma)}{N(x_\text{t}, \sigma)-\frac{1}{2}} \right).
\end{equation}

For $\nicefrac{x_\text{t}}{\sigma} \gg 1$, the particle does not yet
experience the confinement. In this situation
$n(x_\text{t},\sigma) \approx 0$ and $\left<x^2\right> = \sigma^2$. For
$\nicefrac{x_\text{t}}{\sigma} \ll 1$ the chance of finding the
particle in the container is uniformly distributed ($n_\text{t}=\nicefrac{1}{2x_\text{t}})$, and
$\left<x^2\right>$ saturates at $\nicefrac{x_\text{t}^2}{3}$.

When moving to three dimensions, the average squared displacement of the separate dimensions can be simply summed because they are orthogonal.

The diffusion coefficient can only be
determined if there has been a sufficient amount of collisions.
In between the collisions, particles have constant velocity and direction. 
Due to the stochastic nature of the collision events,
the velocity autocorrelation decays exponentially with time
constant~\cite{Langevin1908,Lemons1997}

\begin{equation}
  \label{eq:time_const}
  \tau_\text{v} = \frac{m^*}{f},
\end{equation}

where $f$ [\si{\kilo\gram\per\second}] is the drag coefficient and $m^*$ [\si{\kilo\gram}] is the
effective mass. 

The situation for $t \ll \tau_\text{v}$ is referred to
as the ballistic regime. Here, the average squared distance travelled $\left< x^2 \right>$ is quadratic rather than linear in time. The transition from the (quadratic) ballistic regime to the (linear)
diffusion regime (eq.~\ref{eq:diffusion}) is modelled
phenomenologically by:

\begin{equation}
  \sigma^2 = 6 D \frac{t^2}{t+\tau_\text{v}}.
  \label{eq:std_av_sq_disp}
\end{equation}

Note that both the effective mass $m^*$ and the drag coefficient $f$ depend on the environment. 
The effective mass takes into account the fact that when the
particle is accelerated, the surrounding water mass is also accelerated.
For incompressible fluids with either zero viscosity or infinite
viscosity (Stokes flow), the added mass is \SI{50}{\percent} of the
mass of the water displaced by the sphere~\cite{Landau1987}. For
turbulent flow, both experiment~\cite{Pantaleone2011} as well as
numerical simulations~\cite{Chang1994,Chang1995} show that the added
mass is also to a good approximation \SI{50}{\percent},
irrespective of the Reynolds number or acceleration. There are reports
that the added mass might be bigger in cases where the sphere is
traveling through its own wake~\cite{Odar1964}, which is rare in our experimental setup.
Therefore, we have suggested a simple estimate of the added mass,

\begin{equation}
  \label{eq:madd}
  m^* = m + \textstyle{\frac{2}{3}} \pi r^3 \rho_\text{fluid},
\end{equation}

for a particle with radius $r$ [\si{\meter}] and mass $m$ [\si{\kilo\gram}] surrounded by a
fluid with density $\rho_\text{fluid}$ [\si{\kilo\gram\per\cubic\meter}]. 

\subsection{Velocity distribution}
Li \emph{et al.}~\cite{Li2010b} have experimentally proven that the velocity of particles
performing a Brownian motion is M-B distributed. This distribution of velocity $v$ [\si{\meter\per\second
}] is determined by its mode $v_\text{p}$,

\begin{equation}
  \label{eq:vdistr}
  p(v) = \frac{4 v^2}{\sqrt{\pi} v_\text{p}^3} 
  e^{-\left(\frac{v}{v_\text{p}}\right)^2}.
\end{equation}

At the mode, the distribution reaches its maximum; thus $v_\text{p}$ is the
most probable velocity. For completeness, we note that the average squared velocity is $\left<v^2\right>=\textstyle\frac{3}{2}v_\text{p}^2$.

\subsection{Drag coefficient}
Brownian motion is primarily studied on the microscopic scale,
where the Reynolds number is much smaller than unity. In this case, the
drag force is linear in velocity and the relevant drag coefficient
$f$ is equal to the Stokes drag coefficient. However, on a macroscopic scale, we
deal with turbulent flow and a high Reynolds number, where the drag
force $F_\text{d}$ [\si{\newton}] is quadratic in velocity,

\begin{equation}
  \label{eq:1}
  F_\text{d}=\textstyle{\frac{1}{2}}\rho_\text{fluid}C_\text{d}Av^2,
\end{equation}

where $C_\text{d}$ is the drag coefficient and $A$ [\si{\square\meter}] is the cross
sectional area of the object in the direction of motion.

In our experiment, the particles are continuously ``falling'' through the
upward water flow. This upward flow is set to the terminal
velocity $v_\text{t}$ of the particles, so that they levitate in front
of the camera. Assuming that the changes in the velocity of the
particle caused by turbulence are much smaller than the terminal
velocity, we can obtain an effective drag coefficient by linearising
around the terminal velocity

\begin{equation}
  f = \frac{\text{d} F_\text{d}}{\text{d} v} \Bigr|_{v=v_\text{t}}
  = \rho_\text{fluid} C_\text{d} A v_\text{t}.
  \label{eq:friction}
\end{equation}

\subsection{Disturbing energy}
On the microscale, the diffusion coefficient and velocity distribution
of particles in the fluid can be linked to the temperature.  This
concept can be extrapolated to macroscale systems where disorder is
achieved by shaking rather than by temperature. In that case, one speaks
about effective temperature~\cite{Ilievski2011, Wang2011a}, which is usually
significantly higher than the environmental temperature. Since shaking
can be highly directional, we prefer to
characterize the shaking action by energy ($kT$ [J]) rather than
temperature to avoid confusion.

Starting from the velocity distribution (eq.~\ref{eq:vdistr}), and
considering that $\left<v^2\right>=\nicefrac{3kT}{m}$ for three-dimensional
random walks, the most probable velocity is related to
the kinetic energy through:

\begin{equation}
  \label{eq:kinetics}
  kT = \textstyle{\frac{1}{2}} m^* v_\text{p}^2.
\end{equation}

The Einstein relation also relates the diffusion constant and viscous drag
coefficient of a particle to the thermal energy $kT$:

\begin{equation}
  \label{eq:Einstein}
  kT = fD.
\end{equation}

If particles in a self-assembly reactor behave according to Brownian
motion, both relation~(\ref{eq:kinetics}) and~(\ref{eq:Einstein}) can be
used to obtain the disturbing energy and should give identical results.

\begin{figure}
  \begin{center}
    \includegraphics[width=\narrowfigurewidth]{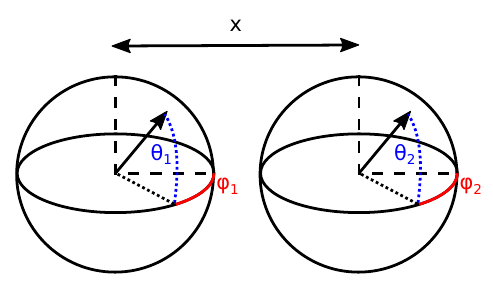}
    \caption{The interaction between two spheres modelled by
      magnetic dipoles at distance $x$ with orientation vector
      $\bm{\theta} = \left[ \theta_1 \ \phi_1 \
      \theta_2 \ \phi_2 \right]$.}
    \label{fig:twoparticles}
  \end{center}
\end{figure}

In addition to measuring the disturbing energy $kT$ from Brownian
motion, we can also estimate it from the interaction between two
attracting magnetic objects. In this situation, we use the fact that he
probability of the system being in a state is governed by
M-B statistics. Consider a system of two spherical
magnetic particles in a confined space
(Figure~\ref{fig:twoparticles}). The chance that the distance of those
particles measured from center-to-center is smaller than $x_0$ is:

\begin{align}
  p(x \leq x_0) =& \frac{1}{Z} \int_d^{x_0} \int_{\boldsymbol{\theta}}  
                    g_\text{r}(x) e^{-\frac{E_\text{m}(\boldsymbol{\theta},x)}{kT}} 
                    d\boldsymbol{\theta} dx \label{eq:MBstatistics} \\
  Z =& \int_d^{D} \int_{\boldsymbol{\theta}} g_\text{r}(x)
        e^{-\frac{E_\text{m}(\boldsymbol{\theta},x)}{kT}} 
        d\boldsymbol{\theta}  dx \nonumber\\
  	\boldsymbol{\theta} =& 
                    \left[ \theta_1 \ \phi_1 \ \theta_2 \ \phi_2 \right]. \nonumber
\end{align}

Here $g_\text{r}(x)$ is the probability density function of a sphere pair with distance 
$x$ between their centers, unaffected by magnetic forces, which models the
influence of the geometry of the reactor.

The distance between the cylindrical magnets is at all times at least a factor of four 
of the magnet height $h$ ($h \leq d/4$). At this point, we approximate their magnetic 
field as well as their magnetic moments by point dipoles. This approximation is accurate 
within \SI{1.3}{\percent} for our magnet geometry. In that case, the magnetic
energy of particle 1 with magnetic moment
$\boldsymbol{m}(\theta_1, \phi_1)$ [\si{\ampere\square\meter}] in a field
$\boldsymbol{B}(\theta_2, \phi_2, x)$ [\si{\tesla}] generated by particle 2
reduces to

\begin{equation}	
  E_\text{m}(\boldsymbol{\theta}, x) =
  -\boldsymbol{m}(\theta_1, \phi_1) \cdot \boldsymbol{B}(\theta_2, \phi_2, x).
\end{equation}

Equation~\eqref{eq:MBstatistics} can be approximated
numerically by a Monte-Carlo approach in which a large number of random combinations of
sphere locations and orientations are selected, yielding different values for  $E_\text{m}$.
The geometry
factor $g_\text{r}$ is approximated by repeated random sampling of two
point locations in a confined
geometry and then gathering statistics about their distance. 


\section{Materials and methods}
\label{sec:materials}

\subsection{Reactor}
\begin{figure}
  \begin{center}
    \includegraphics[width=\linewidth]{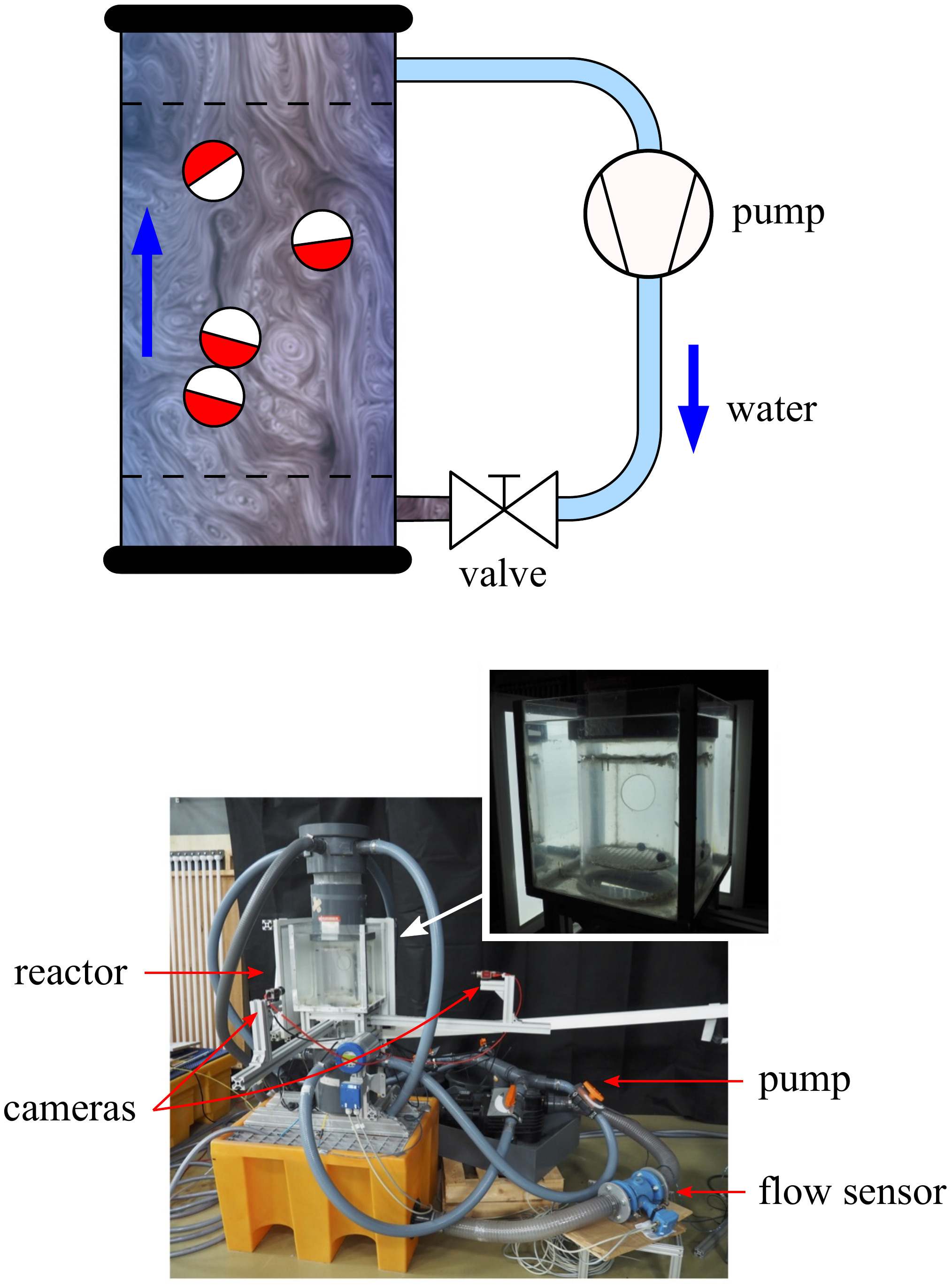}
    \caption{Schematic (top) and experimental (bottom) setup of the
      macroscopic self-assembly reactor. Water is pumped from the
      bottom to the top of the reactor, counteracting gravity and
      supplying energy to the particles via turbulent flow. 
      Meshes prevent the particles from moving outside of the
      field of view of cameras placed around the reactor.}
    \label{fig:reactor}
  \end{center}
\end{figure}

The experimental setup consists of a transparent cylinder
with an inner diameter of \SI{17.3+-0.1}{cm} containing the particles
of interest (Figure \ref{fig:reactor}). Gravity is counteracted by
pumping water from the bottom to the top via four \SI{4.0+-0.1}{cm} diameter
inlet holes using a MAXI.2 40T pump (PSH pools). The water exiting the cylinder is collected in an
open container connected to the pump inlet. 
 The
water flow entering the pump is monitored using an altometer (IFS 4000, Krohne Messtechnik GmbH).

Meshes spaced at \SI{17}{cm} prevent the particles from moving outside
the field of view of cameras placed around the reactor. The dynamics of the particle-fluid system are determined by the particle
density and geometry, as well as water flow speed.

\subsection{Particles}
The particles used in the experiments are 3D-printed polymeric (ABS)
spheres with a diameter of \SIrange{1.67 \pm  0.01}{2.02 \pm 0.02}{\centi\meter} 
and a
corresponding density of \SIrange{1.33 \pm 0.02}{1.25\pm 0.04}{\gram\per\cubic\centi\meter} (larger
particles have lower density).
The core of the spheres consist of
cylindrical, axially magnetised NdFeB magnets with a length of
\SI{3.8+-0.05}{\milli\meter} and a diameter of \SI{3.8+-0.05}{\milli\meter} (Supermagnete, grade N42, Webcraft GmbH). 
The dipole moment (\SI{50.8(1)}{\milli\ampere\square\meter}) was determined by 
measuring the force between two magnets using a balance.
 
\begin{figure}
  \begin{center}
    \includegraphics[width=\linewidth]{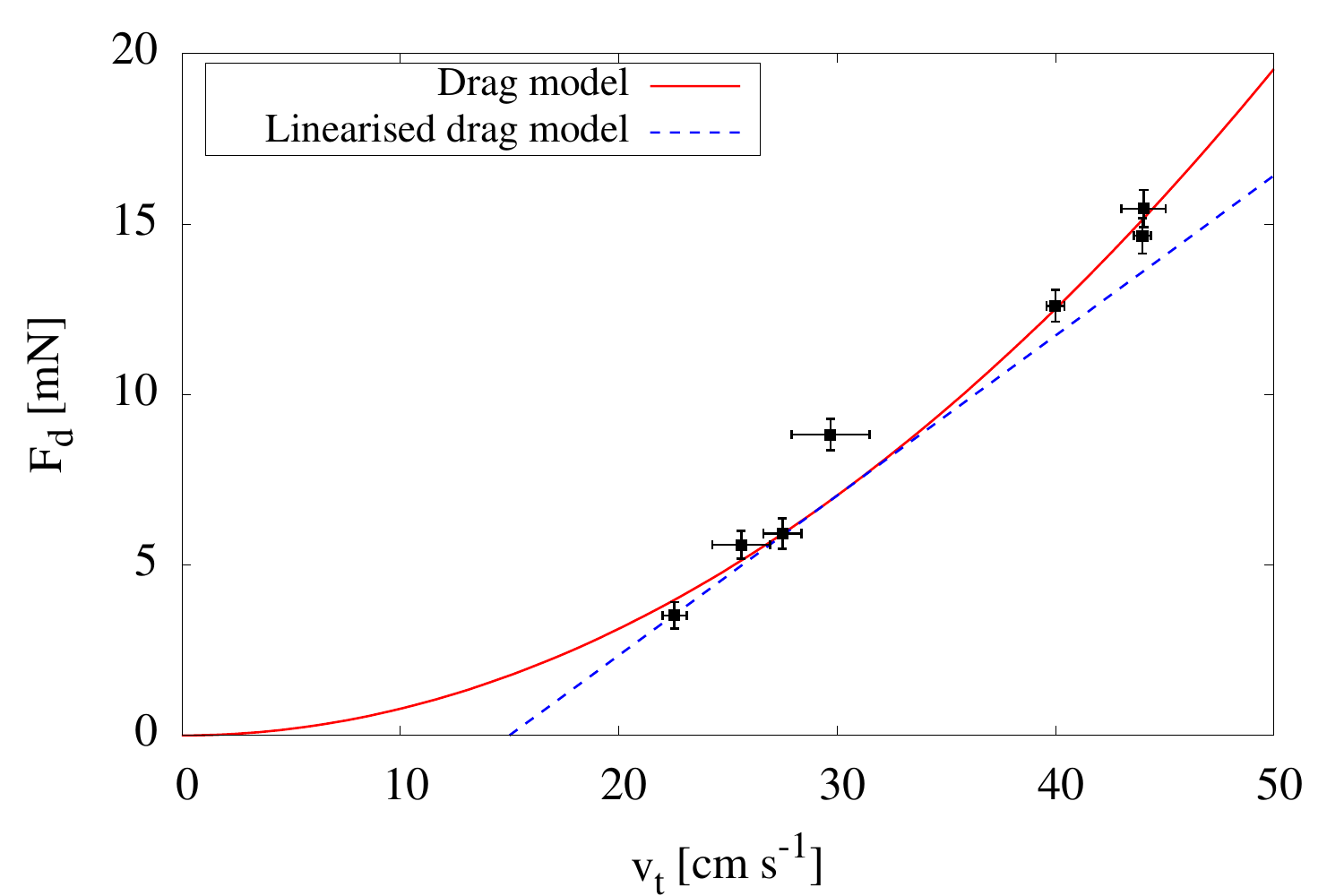}
    \caption{Calculated drag force versus measured terminal velocity for spheres
      with equal diameter but varying densities. The effective drag
      coefficient is obtained by linearisation around the terminal
      velocity (eq.~\ref{eq:friction}), illustrated by the blue
      dashed line for $v_\text{t}$=\SI{30}{\centi\meter\per\second}.}
    \label{fig:drag}
  \end{center}
\end{figure}

The drag coefficient of the particles was estimated from their
terminal drop velocity. For this, particles with a range of densities but identical 
diameter of \SI{1.85}{\centi\meter}
were released at the
top of a \SI{2}{\meter} high cylinder filled with water. Once an equilibrium 
between drag- and gravitational force had been established (approximately \SI{0.5}{\meter} after release), the velocity of
the particles was measured with a video camera over a distance of
\SI{1.0}{\meter}.
Figure \ref{fig:drag} shows the measured relation between drag force
and terminal velocity. From fitting equation \ref{eq:1}, we obtain 
$\frac{1}{2}\rho_\text{fluid}C_\text{d}A =$
\SI{78(3)}{\gram\per\meter}.  Assuming the density of water to be
\SI{1000}{\kilo\gram\per\cubic\meter}, we obtain
$C_\text{d}$=\num{0.58+-0.02}. Spheres of this diameter and velocity
in water have a Reynolds number of approximately \num{5500}. At this
value, Brown \emph{et al.}~\cite{Brown2003} predict
$C_\text{d}$=\num{0.39}, which is substantially lower. The reason for
the discrepancy is unknown to us. The measured drag coefficient is
used in the remainder of this paper.



\subsection{Reconstruction}
Two calibrated, synchronised cameras (Mako G-131, Allied Vision) were placed around the 
reactor at an angle of approximately \SI{90}{\degree} and they recorded datasets at 
\SI{30}{fps} at a resolution of \num{640}$\times$\num{512}. The reactor is 
surrounded by a square, water-filled aquarium to prevent refraction due to its 
cylindrical nature. Backlight panels were used to enhance 
contrast. Single spheres were observed for \SI{15}{min} and two spheres for 
\SI{30}{min}. Offline, the location of the spheres was automatically detected 
using a custom written \textsc{matlab} script. 
A method based on the direct linear 
transform algorithm~\cite{Hartley2004} was used for 3D reconstruction, giving an average 
reconstruction error of \SI{0.16}{\centi\meter}. Trajectories 
closer than \SI{1.5}{cm} to the meshes were discarded to rule out the 
significant effect of the altered hydrodynamic interaction at these interfaces.
The velocity vector of the particle is obtained by $\mathbf{v} = \Delta 
\mathbf{x} f_\text{cam}$, the product of the particle 
displacement between two frames, and the camera frame rate. 

\section{Results}
\label{sec:results}

\subsection{Single particles}

\begin{figure}[tb]
	\begin{center}
			\includegraphics[width=\narrowfigurewidth]{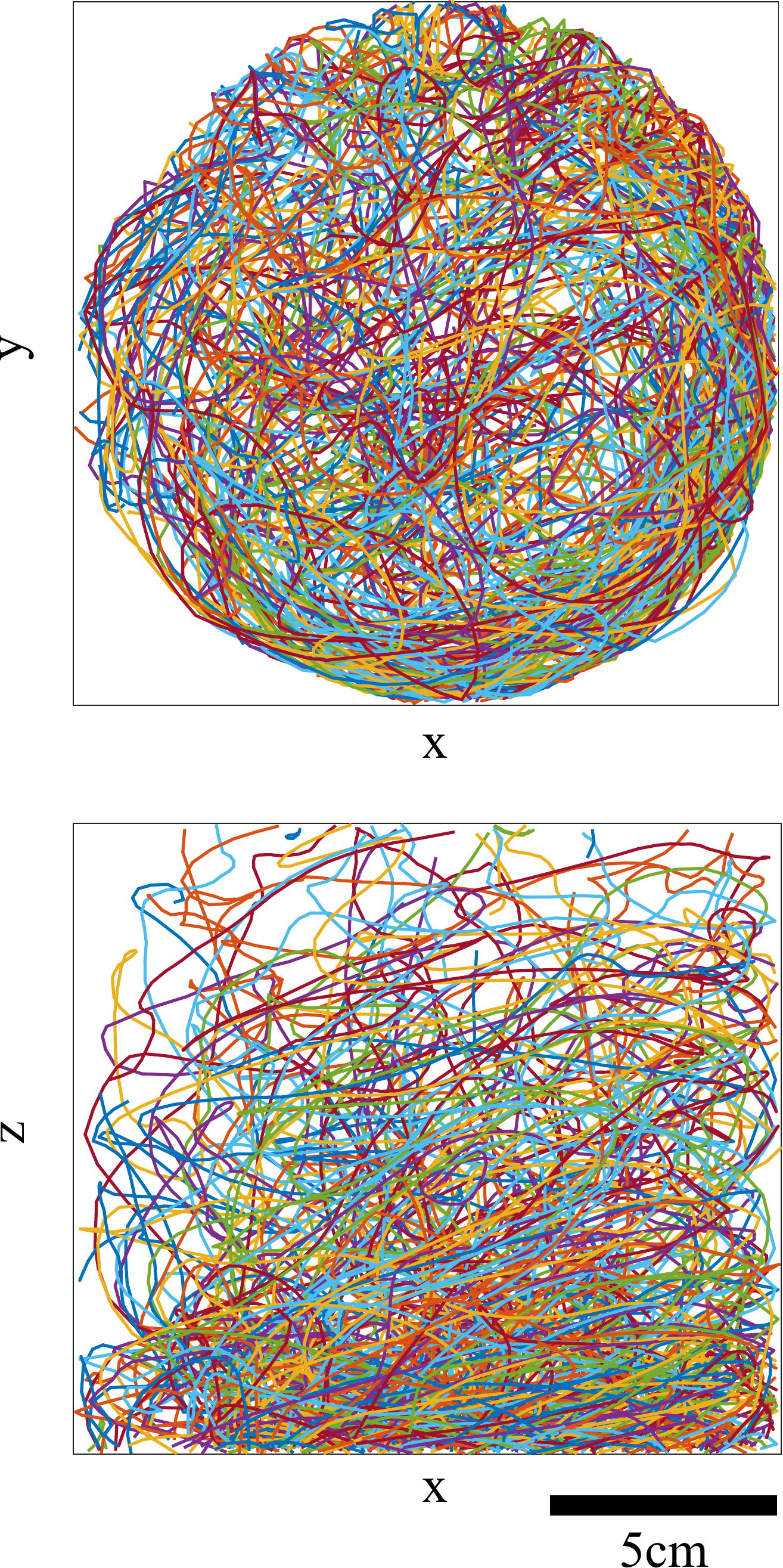}
		\caption{Top (upper) and side (bottom) view of the reconstructed trajectories 
of a single sphere (diameter \SI{1.85}{\centi\meter}) moving through the 
reactor. Coordinates less than \SI{1.5}{\centi\meter} close to the top- and 
bottom meshes are removed to rule out significant influence of the meshes. In this way, 
the single trajectory is cut into many smaller ones, which are each 
assigned a different color. See Supplementary Material at [URL will be inserted by publisher]
for a three dimensional representation of the data.}
		\label{fig:trajectories}
	\end{center}
\end{figure}

Figure \ref{fig:trajectories} shows a set of reconstructed
trajectories of a \SI{1.85}{cm} sphere in the reactor. Each trajectory
starts and ends when exiting and entering the areas within \SI{1.5}{\centi\meter} of the meshes, and is indicated by a different color.

\begin{figure}[tb]
  \begin{center}    \includegraphics[width=\linewidth]
    {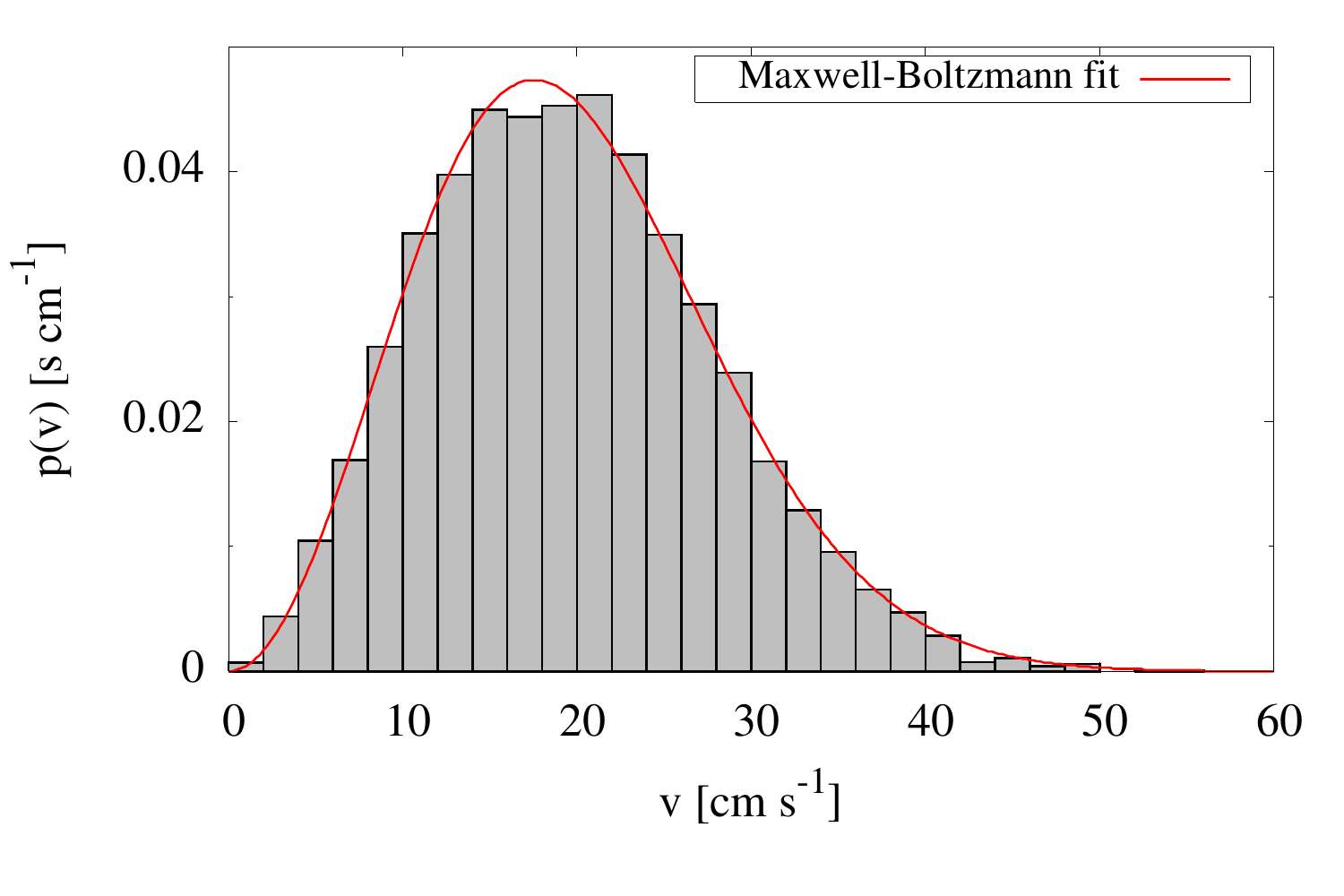}
    \caption{Maxwell-Boltzmann (M-B) distribution fitted to the measured
      velocity distribution of a particle with a diameter of
      \SI{1.80}{\centi\meter}. The Kolmogorov-Smirnov (K-S) test quantifies a maximum 
      distance between the theoretical and experimental
      cumulative distributions of \SI{0.0073} with a Q value of
      \SI{0.70}, indicating a high probability that the velocity is
      indeed M-B distributed.}
    \label{fig:velocity}
  \end{center}
\end{figure}

Figure \ref{fig:velocity} shows the velocity calculated from these trajectories. The histogram is
obtained from the absolute velocity (\num{10600} data points) of a \SI{1.80}{\centi\meter}
sphere. A M-B distribution was fitted to the data by minimising the maximum distance $E_\text{max}$ between
the cumulative empirical and cumulative M-B distribution, yielding fitting parameter $v_\text{p}$. A 
Kolmogorov-Smirnoff (K-S) test was used to quantify the quality of fit and to obtain a 
significance level Q to disproof the null hypothesis that the two distributions are the same.~\cite{Press1992}
With a $E_\text{max}$ of \num{0.0073} and a Q of \num{0.70}, we have good reason to assume that the
velocity is MB-distributed.

\begin{figure}[tb]
  \begin{center}
    \includegraphics[width=\linewidth]
    {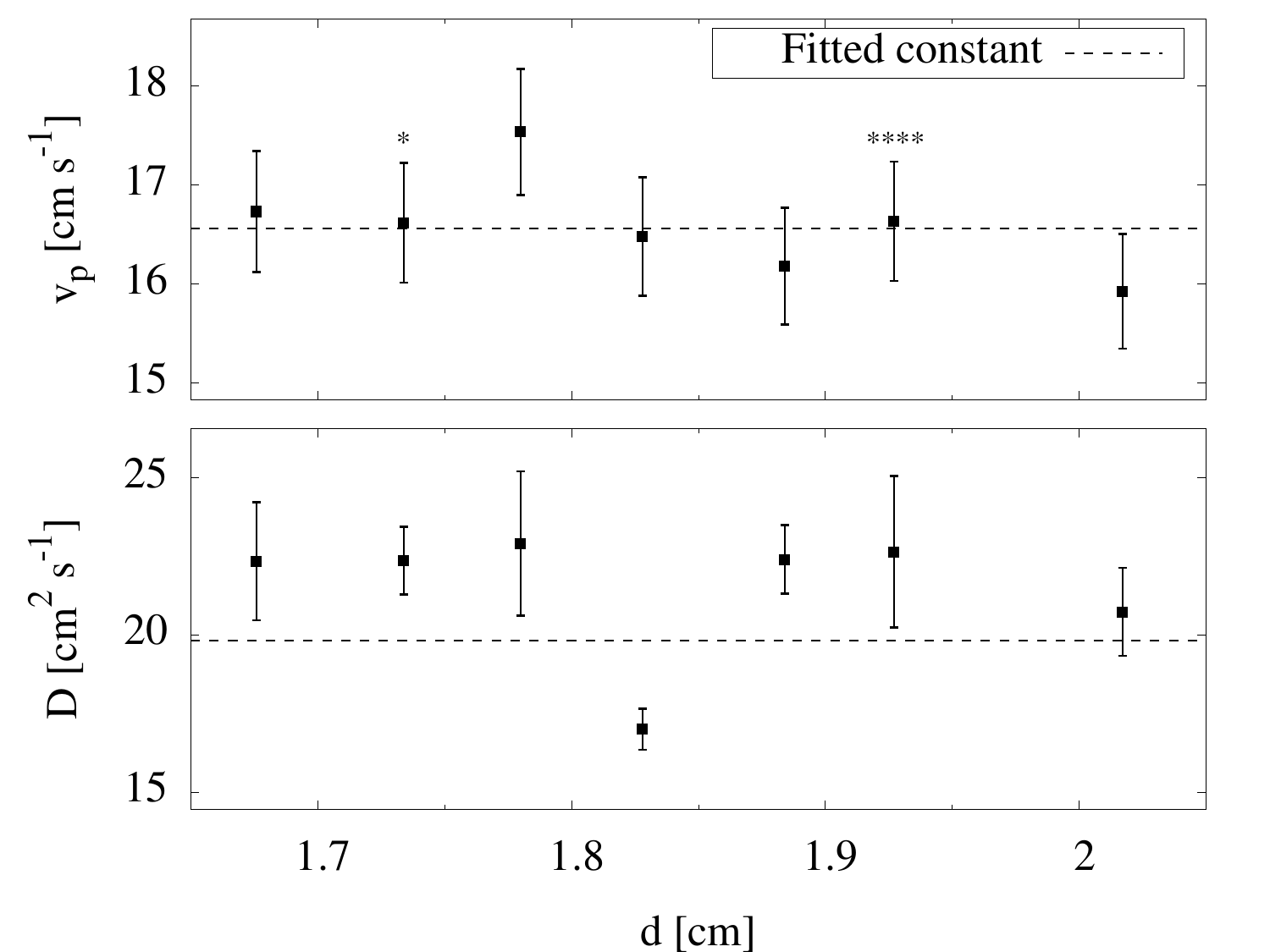}
    \caption{\textbf{Top:} Mode of the M-B distribution
      obtained by fitting to the measured velocity distribution of
      particles of various diameters (reduced $\chi^2=0.68$,
      $Q=0.67$). Stars indicate the quality of fit (Q-value) of the 
      K-S test (*~$<0.05$, ****~$<0.0001$).
      \textbf{Bottom:} Diffusion coefficient obtained by
      fitting the diffusion model to the average square displacement
      (reduced $\chi^2=5.85$, $Q=4\cdot10^{-6}$).}
    \label{fig:param_fit}
  \end{center}
\end{figure}

Figure \ref{fig:param_fit} displays the resulting $v_\text{p}$ for
spheres of various diameters, for which we find a range from
\SIrange{15.92}{17.54}{\centi\meter\per\second}. The fit to the
M-B distribution has a $Q$-value above \num{0.05} for five
out of the seven measurements. Even though the data suggests a slight
decrease of velocity with increasing sphere size, the particle
velocity fits very well to a model assuming constant velocity, with an
average of 
\SI{16.6(2)}{\centi\meter\per\second}. This analysis was carried out
using a chi-square fitting routing, yielding the reduced $\chi^2$ error metric
(ideally being around \num{1}) and the corresponding Q-value (the probability
that a $\chi^2$ equal or greater than the observed value is caused by chance).~\cite{Press1992}
The reduced $\chi^2$ of this fit is close to
unity (\num{0.68}) with a very high $Q$-value of \num{0.67}.

\begin{figure}[tb]
  \begin{center}
    \includegraphics[width=\narrowfigurewidth]
    {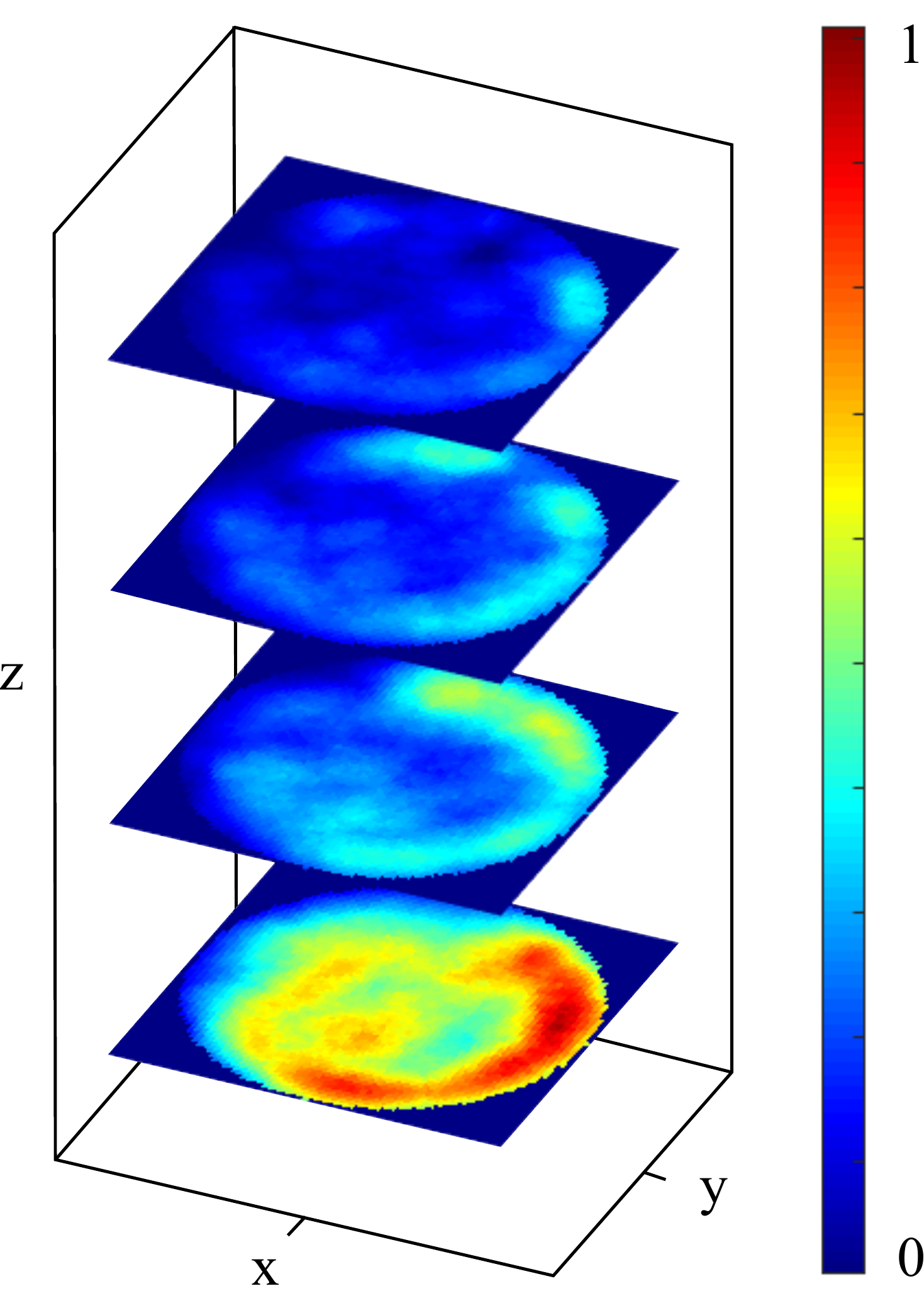}
    \caption{Normalised probability distribution of a single sphere
      (diameter \SI{1.85}{\centi\meter}) in the reactor, displayed in
      slices along the reactor tube. The particle has a clear
      preference for the bottom region as well as the edge
      regions. Quantised, the particle has a chance of
      \SI{62}{\percent}, \SI{48}{\percent} and \SI{26}{\percent} to be
      in, respectively, the right (positive $x$-coordinate), back
      (positive $y$-coordinate) and top (positive $z$-coordinate) halves
      of the reactor.}
    \label{fig:sphere_prob}
  \end{center}
\end{figure}

Figure \ref{fig:sphere_prob} shows the normalised distribution of the particle
at several $z$-slices across the reactor. It can be seen that the
particle has a preference for the bottom area, especially near
the reactor walls of the positive $x$-coordinate. We believe that this
phenomenon is caused by a non-uniform flow pattern of water that results
from the specific valve settings.

\begin{figure}
	\begin{center}
		\includegraphics[width=\linewidth]{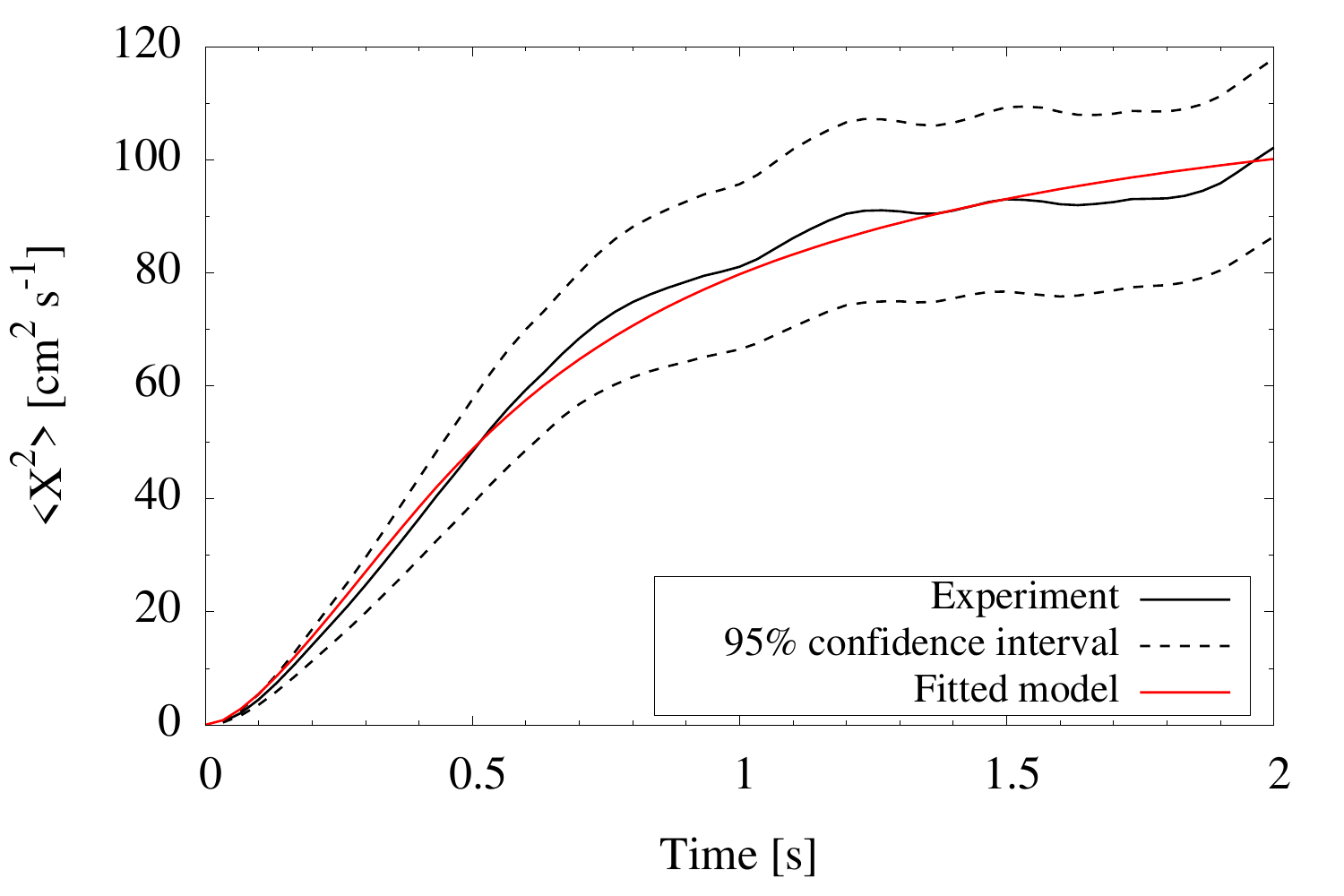}
		\caption{Average squared displacement as a function of time for a sphere with 
diameter \SI{1.90}{cm}, calculated from 65 trajectories. The model fits  within 
the \SI{95}{\percent} confidence interval.}
		\label{fig:diffusion}
	\end{center}
\end{figure}

The average squared displacement was
calculated from the longest trajectories; that is, those with a minimum duration of
\SI{2.0}{\second}. Figure \ref{fig:diffusion} shows the resulting curve for
a sphere with a diameter of \SI{1.90}{\centi\meter}. The curve shows a 
quadratic regime below \SI{0.3}{\second}, shortly entering an approximate linear 
regime before slowly converging to a horizontal asymptote.

The movement of the sphere is in the quadratic, or ballistic, regime
when the measurement time is shorter than the average time between
directional changes (``collisions''), 
$\tau_\text{v}$. Using
  measured values for the drag coefficient and effective mass in
  equation~\ref{eq:time_const}, we obtain values for $\tau_\text{v}$
  ranging from \SIrange{134}{149(10)}{\milli\second}. The saturation measured for longer observations is caused by the confined geometry of the reactor and it will change as the
reactor is changed in shape and size.

The model described by equations \ref{eq:av_sq_disp_analytic} and \ref{eq:std_av_sq_disp}
was fitted to the measurements, yielding values for diffusion coefficient $D$ and
average reactor size $x_\text{t}$.

We have to take into account that the model has its limitations.
First of all it is based on a symmetrical truncated normal
distribution. This would require the particle to always start in the
center of the reactor. In contrast, all of the measured
trajectories start at a random place at the top or bottom of the
reactor due to the method that we used to obtain separate trajectories.

Secondly, the cylindrical geometry of the reactor is not included 
in the model. These two issues mainly affect the estimation of the 
reactor size. 

Finally, the ballistic regime was phenomenologically modelled 
without physical background. This region, which has 
a high weight factor during fitting the model to the data (due to the small 
error bars in the data), can result in a significant fitting error.

Given that only the latter aspect could give errors in the estimation of $D$, we consider 
the obtained values for $D$ to be quite reasonable, with values between \num{17}
and \SI{23}{\square\centi\meter\per\second} (see figure \ref{fig:param_fit}). 
The average diffusion coefficient for all of the measured diameters is
\SI{20(1)}{\square\centi\meter\per\second}. Judging from the
graph, there seems to be no reason to assume that the diffusion
coefficient has a strong dependence on sphere diameter. It should be
noted, however, that this assumption leads to a very high reduced
$\chi^2$ (\num{5.85}) and low quality of fit
$Q$ ($4\cdot10^{-6}$). However, due to the previously mentioned model inaccuracies, 
we think that we may have underestimated the errors in the estimation of $D$.


\subsection{Two-sphere results}

\begin{figure}
	\begin{center}
\includegraphics[width=\linewidth]
{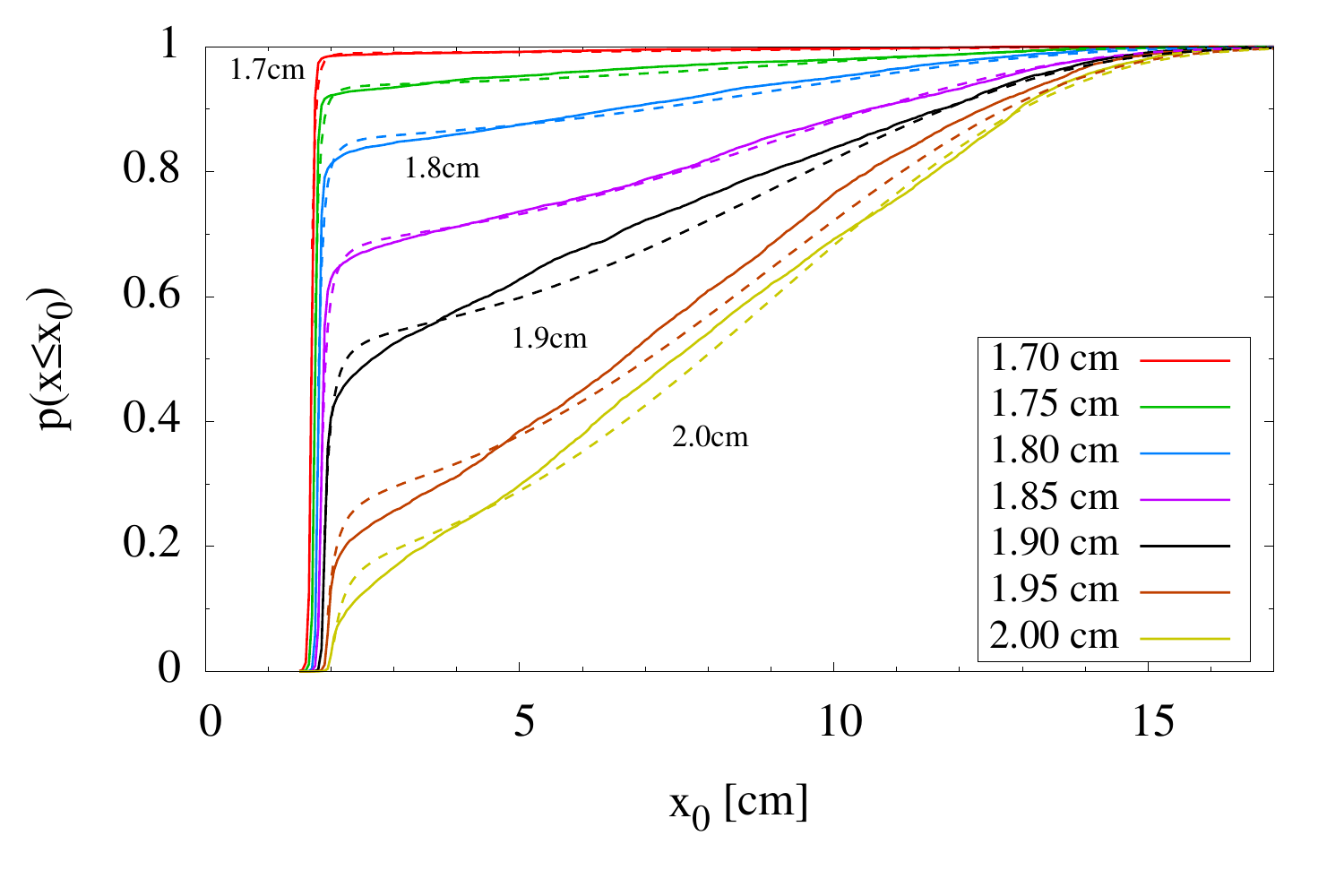}
\caption{Measured probability (cumulative) of the distance between the
  centers of two magnetic spheres ($x$) for various sphere
  diameters. A model based on M-B statistics captures
  the shapes of the curves with a maximum error of \SI{5}{\percent} of
  the full range. As the spheres decrease in size, they are more likely to
  be in a connected state.}
		\label{fig:twoParticles}
	\end{center}
\end{figure}

From the two-sphere experiments, the distance $x$ between the particles was 
tracked over time. Figure~\ref{fig:twoParticles} shows the cumulative 
probability of sphere distance $p(x \leq x_0)$ for 
spheres of various diameters. Spheres with smaller diameters have a lower 
magnetic energy in connected state and, therefore, a higher probability of being 
connected. In other words, $p(x \leq d)$ becomes larger for smaller $d$. 
All of our measurements follow a similar profile: they consist of a curved regime 
for $x\leq \SI{3}{\centi\meter}$ followed by an approximately linear region 
for $x>\SI{3}{\centi\meter}$. The linear regime indicates that
magnetic forces are no longer significant for particle interaction. For 
$x>\SI{13}{\centi\meter}$ there is a saturation effect caused by the reactor 
geometry.
The model of equation~\ref{eq:MBstatistics} has been fitted to the curves by
minimising the maximum distance between the curves (based on the Kolmogorov-Smirnoff
method~\cite{Press1992}). Although this is not an exact fit, it manages to 
capture the shape with a maximum error of \SI{5}{\percent} of the full range.

\subsection{Disturbing energy}

The experiments provide three methods for the characterisation of the
equivalent thermal energy of the system. Numerical values for the
kinetic energy were calculated from the measured velocity and added
mass according to equation~\ref{eq:kinetics}. The measured diffusion
coefficient and drag coefficient at the set water flow speed (equation~\ref{eq:friction})
were used to calculate the energy using the Einstein relation (equation~\ref{eq:Einstein}).
Additionally, two-particle experiments provide numerical values for
the equivalent energy as a result of fitting equation~\ref{eq:MBstatistics}
to the measured data, as depicted in figure~\ref{fig:twoParticles}.

\begin{figure}
  \begin{center}		
    \includegraphics[width=\linewidth]
    {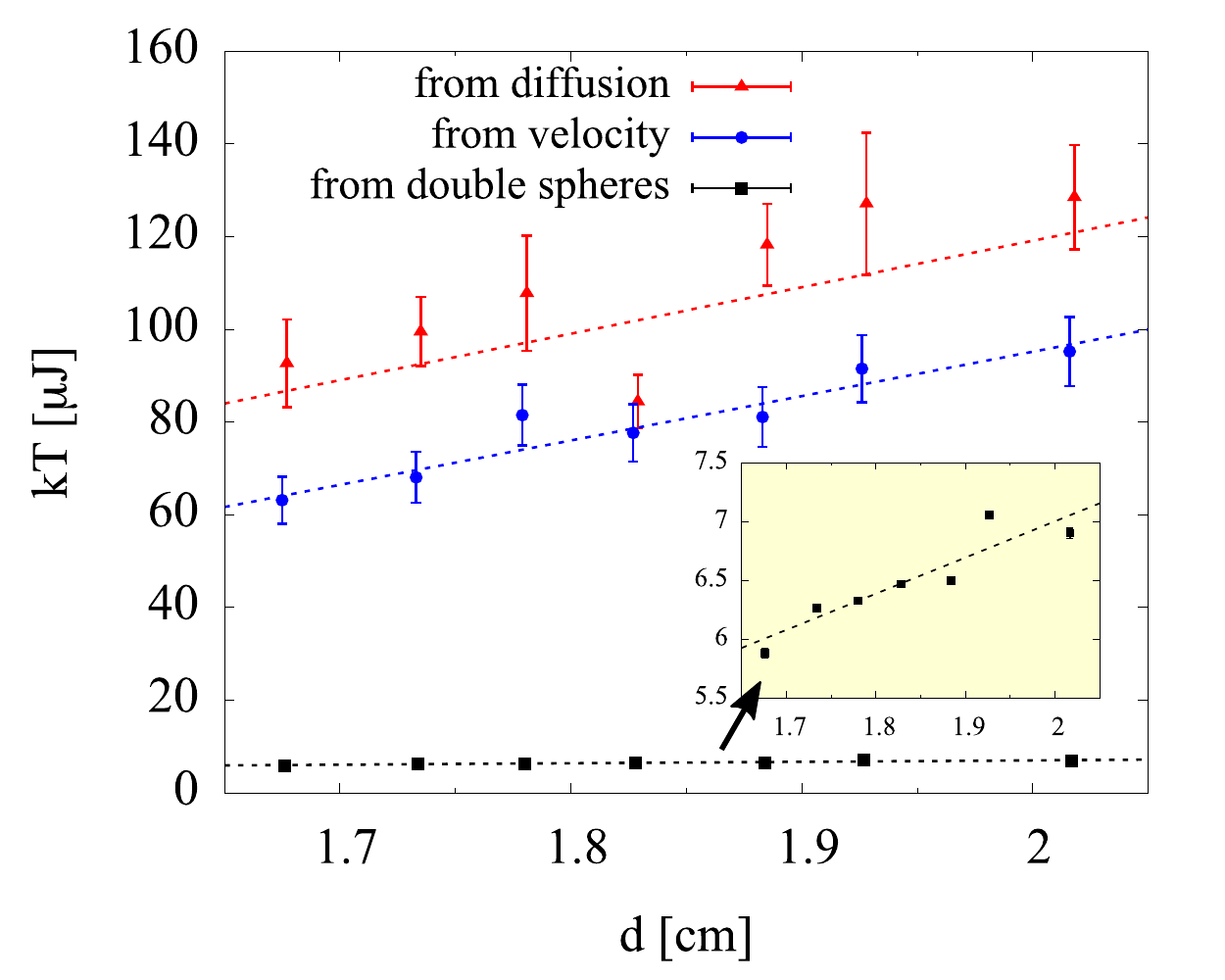}
    \caption{Disturbing energy of the turbulent field calculated from
      the diffusion coefficient, the velocity distribution and double
      sphere experiments.  The disturbing energy estimated from the
      single sphere experiments (diffusion, velocity) are
      approximately a factor 10 higher than that estimated from double
      sphere experiments. The dashed lines are guides to the eye. There is
      an increase in energy with an increase in sphere diameter, which is proportional with the increase in mass and friction coefficient.}
      \label{fig:energy}
    \end{center}
  \end{figure}

The resulting values for
all of the spheres are summarised in figure~\ref{fig:energy}. A first
observation is that the results obtained via single sphere
experiments (velocity, diffusion) are in the same order of magnitude,
and differ approximately \SI{20}{\micro\joule}. They span a range from 
approximately \SIrange{60}{120}{\micro\joule}.
These values are, however, more than a factor of ten higher than the results
obtained via the two sphere experiments, which range from approximately
\SIrange{6}{7}{\micro\joule}. The possible origin for this discrepancy
is discussed in the following section.

In all cases, the energy increases as the sphere size increases, by approximately 
\SI{17}{\percent}, \SI{41}{\percent}, and \SI{46}{\percent} for, respectively, 
two-sphere experiments, diffusion, and velocity. As we concluded previously, the diffusion 
coefficient and average sphere velocity do not depend on the sphere size 
(figure \ref{fig:param_fit}). The increase of energy is caused by an increase
in mass and friction coefficient, and both are dependent on sphere radius.

\section{Discussion}
\label{sec:discussion}

From the trajectory analysis of single particles, we were able to determine 
that their velocity distribution closely follows a M-B distribution. Additionally, we 
have seen that the average squared displacement as a function of time follows a shape 
that was predicted by a confined random walk model. These conclusions strongly support 
the hypothesis that particles in the reactor perform a random walk.

When increasing the particle size, the observed disturbing energy $kT$
also increases. However, there is no observable increase in velocity or
diffusion coefficient. For the energy calculated via velocity
and diffusion, this means that this increase in energy is caused by an
increase in, respectively, effective particle mass and drag
coefficient. The corresponding curves, as shown in figure
\ref{fig:energy}, are very similar due to the fact that the particle
mass and drag force are coupled. With an increase in particle radius,
both the mass and surface area are
increased. 
The increase in energy occurs without physically changing the nature
of the disturbing energy; that is, the speed and turbulence of the water
flow is unaltered. This means that the amount of energy that is
transferred from the environment to the particle is dependent on the
particle geometry. 

An explanation for this effect might be found in the wavelength
dependence of the turbulence. Turbulence is introduced as a large
wavelength disturbance at the bottom of the cylinder, after which it
propagates upwards in an energy cascade that transfers the energy to
smaller wavelengths. This process is dissipative (Richardson
cascade~\cite{Richardson1922}). The resulting energy spectrum
drops off at increasing wave numbers.~\cite{Kolmogorov1941} 
Therefore, we can assume that the disturbing energy as
experienced by the particles is not, like in Brownian motion,
characterised by a flat spatial frequency spectrum (white noise) but
instead drops off at shorter wavelengths. So, effectively, the bandwidth of the
energy transfer increases for larger particles.

The assumption of a dissipative energy cascade could also explain why
the energy obtained from two-sphere experiments is lower compared to
single sphere experiments. While all of the spatial frequency components in
the turbulent flow drive an object around the system in a random walk,
wavelengths in order of the particle diameter contribute most
effectively to separation of connected particles. The disturbing
energy dropping with decreasing wavelength would explain why the
disturbing energy estimated from the two particle experiment is smaller
than that obtained from the random walk.

It is perhaps in the spatial frequency spectrum where the analogy
between turbulent flow and true Brownian motion breaks down. Therefore, we
will need to characterise the effective energy of the system
separately for particles of different size. Special care needs to be
taken for large clusters of particles because they are effectively a large
particle and, therefore, subject to a higher energy portion. At the same
time, particle-particle interaction is subject to a lesser amount of
disturbing energy. Consequently, such systems will have a bias towards
the occurrence of smaller particle clusters.


\section{Outlook}
Successful self-assembly is characterised by the ability of the system
to end up in a desired end-state, generally the global energy
minimum. This will require an interplay in assembling and disturbing
forces which assist the system by removing itself from local energy
minima. The experimental results have proven that particles in the
reactor show a Brownian-like motion and that the disturbing turbulent
field is able to separate otherwise connected particles. This gives
confidence that multi-particle systems will be able to explore the energy
landscape and that the results have significance for similar processes
taking place on the microscale.

\begin{figure}
  \begin{center}
    \includegraphics[width=\linewidth]{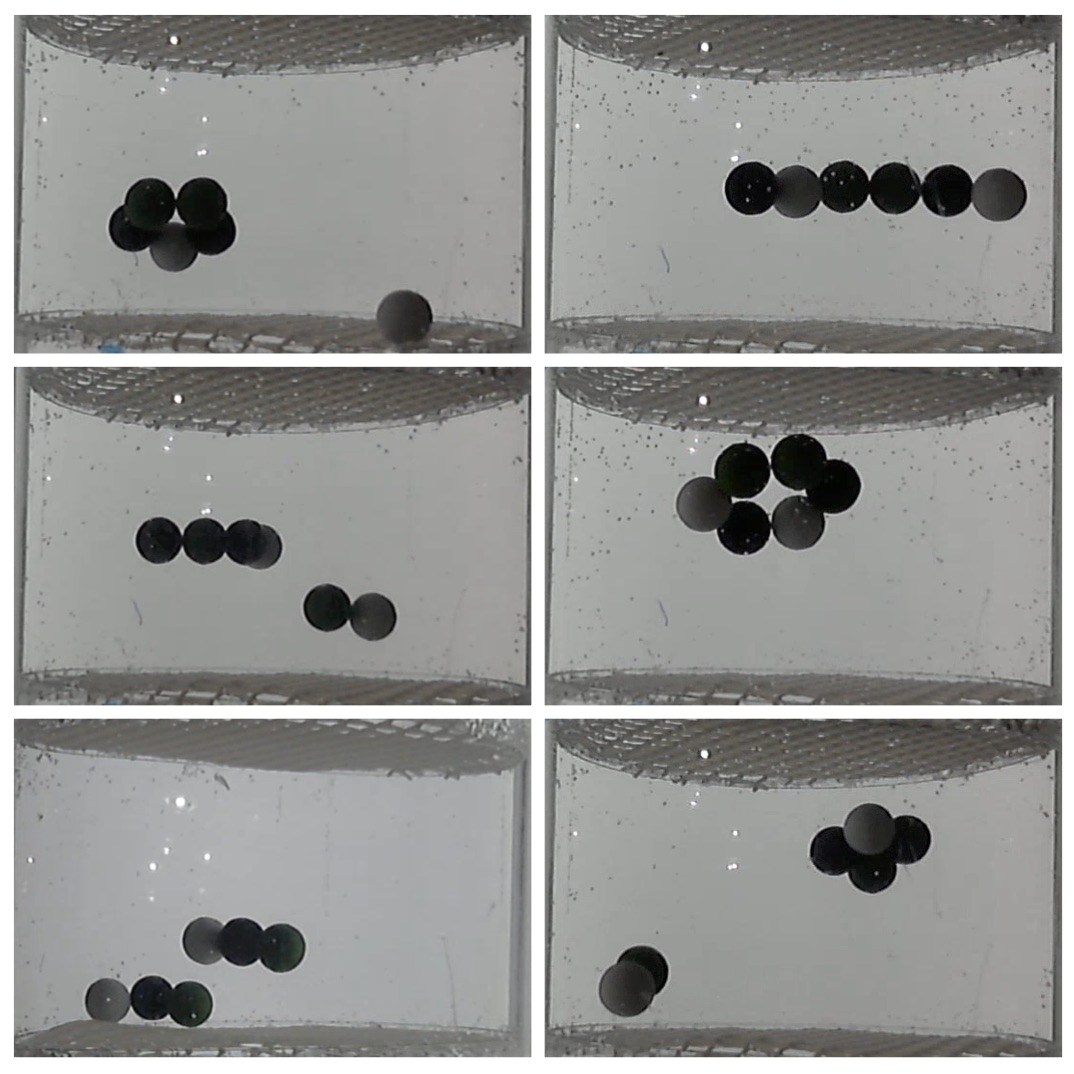}
    \caption{Multi-particle systems show to explore the energy
      landscape, ending up in both local and global energy
      minima. See Supplementary Material at [URL will be inserted by publisher]
	for a video of the processes taking place.}
    \label{fig:sixspheres}
  \end{center}
\end{figure}

To demonstrate the possibilities of using this experimental setup for further studies, we
loaded the reactor with six spheres with embedded magnets. Figure
\ref{fig:sixspheres} shows several stills of a video in which the
spheres form different structures, thereby exploring the energy landscape. The
highest energy state is found when all of the spheres are disconnected. The
energy of the system decreases with the number of connections made, so
a six sphere chain structure (top right) has a lower energy than two
three-sphere structures (bottom left). One more bond can be created
by forming a six-sphere ring (center right). For structures with more
than four spheres, the ring is the minimum energy
state~\cite{Messina2014}. Indeed, three-sphere rings are hardly ever
observed. By long term observation, one could measure the relative
occurrence of the different structures and check if they agree with
Boltzmann statistics.

\section{Conclusions}
We have constructed an experimental setup that allows us to study
the (dis)connection dynamics of centimeter-scale objects by analysing the interaction of
magnetic attraction forces and disturbing turbulent forces. This
``macroscopic self-assembly reactor'' serves as a physical simulator
of self-assembly processes on the microscale and nanoscale, allowing easy
observation by drastically increasing both the length and time
scales.

Trajectory analysis of single spherical particles shows that they
perform a random walk, which analogous to Brownian motion. Spheres with
diameters ranging from \SIrange{1.7}{2.0}{\centi\meter} have a range
of velocities that are M-B distributed.
The most probable velocity (mode) is independent on
sphere size and has a value of
\SI{16.6(2)}{\centi\meter\per\second}. The average square displacement
over time, or the `diffusion profile,' fits to a confined random walk
model.
The diffusion coefficient appears to be independent of sphere size, with an average value of
\SI{20(1)}{\square\centi\meter\per\second}. Although statistical analysis 
disproves this statement, we believe that the measurement error has been 
underestimated.

The particle distribution is non-uniform over the reactor. 
The particle is, for
instance, three times as often in the bottom half of the reaction compared to
the top half. Although this non-uniform distribution does not affect the
Brownian motion behaviour, it virtually reduces
the reactor size.

In two-particles systems, we observe
self-assembly dynamics; that is, the particles occasionally
connect and
disconnect. The cumulative distribution of the distance between the
centers of the particles fits with a maximum error of \SI{5}{\percent} 
of the full range of the distribution to a model based
on M-B statistics.

The disturbing energy (analogue to temperature) of the reactor was
estimated from the velocity distribution and diffusion (single
particle experiments), as well as from the dynamic interaction of
two-particle systems.  The estimates of the disturbing energy
determined from single sphere experiments are in the same order of
magnitude.  However, the disturbing energy obtained from two-sphere
experiments is at least one order of magnitude lower (approximately
\SI{6.5}{\micro\joule} compared to \SI{80}{\micro\joule}). From this
we can conclude that for self-assembly studies, the disturbing energy
of the system cannot be calibrated from single sphere experiments
alone.

  The disturbing energy increases with increasing sphere diameter, from
  \SIrange{1.7}{2.0}{cm}. For the single sphere experiment, this
  increase is more prominent (\SI{41}{\percent} via diffusion analysis, 
  \SI{46}{\percent} via velocity analysis) than for the two-sphere 
  experiment (\SI{17}{\percent}). We reason that the energy
  transfer from the turbulent environment to the particles is dependent
  on particle size and geometry.

In addition to the two-sphere experiment, periodic connection and
disconnection events have also been observed for a six-sphere system, forming ring- and line-based structures. This
demonstrates that the reactor can be successfully applied to study
self-assembly processes at convenient length and time scales, and it may be a 
good simulator for microscopic environments.


\section*{Acknowledgements}
The authors would like to thank Remco Sanders for constructing the
self-assembly reactor and L\'eon Woldering for initial work on the
project. Additionally, we thank John Sader for introducing the concept
of added mass and Marc Pichel for his participation in scientific
discussions and general support.  We also recognise the valuable
contribution of Nikodem Bienia, Donghoon Kim, Gayoung Kim, Yannick
Klein and Minyoung Kim to the scientific work.  Finally, we would like
to thank Bronkhorst BV for providing the flow meter, and Eckard Breuers
for kindly providing nets of various dimensions for our experiments.



\bibliographystyle{apsrev4-1}
%

\end{document}